%% file: main.tex
\documentclass[nobib]{MSword}

\usepackage[english]{babel}
\usepackage{csquotes}
\usepackage[super,numbers,sort&compress]{natbib}
\usepackage{subcaption}
\usepackage{lipsum}
\usepackage[mathscr]{euscript}
\usepackage{tikz}

\usetikzlibrary{shapes.geometric, arrows}
\tikzstyle{startstop} = [rectangle, rounded corners, 
minimum width=3cm, 
minimum height=1cm,
text centered, 
draw=black, 
fill=red!30]

\tikzstyle{io} = [trapezium, 
trapezium stretches=true, 
trapezium left angle=70, 
trapezium right angle=110, 
minimum width=3cm, 
minimum height=1cm, text centered, 
draw=black, fill=blue!30]

\tikzstyle{process} = [rectangle, 
minimum width=3cm, 
minimum height=1cm, 
text centered, 
text width=3cm, 
draw=black, 
fill=orange!30]

\tikzstyle{decision} = [diamond, 
minimum width=3cm, 
minimum height=1cm, 
text centered, 
draw=black, 
fill=green!30]
\tikzstyle{arrow} = [thick,->,>=stealth]
\usepackage[version=3]{mhchem}

\usepackage{tikz,lipsum,lmodern}
\usepackage[most]{tcolorbox}
\usepackage{xfrac,bigints}
\usepackage{authblk}
\usepackage{notoccite}

\usepackage{hyperref}
\hypersetup{
     colorlinks = true,
     citecolor  = blue,  
     linkcolor  = magenta 
}
\usepackage{orcidlink}
\usepackage{color}

\usepackage{makecell}
\usepackage{multirow}
\usepackage{multicol}

\title{ \bf Do 1-dimensional metals prefer to form even-numbered van der Waals clusters ?}

\author[1]{Subhojit Pal\,\orcidlink{0009-0003-6356-3409\,}\thanks{\href{mailto:palsubhojit429@gmail.com}{\dag\ palsubhojit429@gmail.com}}}
\author[2]{John F. Dobson\,\orcidlink{0000-0002-7582-1378\,}\thanks{\href{mailto:john12dobson@gmail.com}{\ddag\ john12dobson@gmail.com}}}
\affil[1]{\emph{\rm Dipartimento di Fisica e Chimica, Emilio Segr\`e, Universit\`a degli Studi di Palermo,
Via Archirafi 36, 
90123 Palermo, Italy}}
\affil[2]{\emph{\rm School of Environment and Science, 
Griffith University, 
Nathan, Queensland 4111, Australia}}
\affil[**]{Corresponding author(s) E-mail(s) : \dag, \ddag}

\date{\today}

\begin{document}

\maketitle
\newpage
\begin{abstract}
Parallel quasi-one-dimensional metals are known  to experience strong
dispersion (van der Waals, vdW) interactions that fall off unusually slowly
with separation  between the metals. Examples include nanotube brushes, nano-wire arrays, and also common biological structures. In a  many-stranded bundle, there are potentially strong multi-strand vdW interactions that go beyond a simple sum  of negative (attractive) pairwise inter-strand energies.
Perturbative analysis  showed that these contributions alternate in sign, with the odd (triplet, quintuplet, ...) terms being positive (repulsive). The triplet case leds to the intriguing speculation that
these strands may prefer to coalesce into even-numbered bundles, which could
have implications for the formation kinetics  of DNA, for example. Here we
use a non-perturbative vdW energy analysis to show that this conjecture is
not true in general. As our counter-example we consider 6 strands and show
that 2 well-separated bundles of 3 strands have a more negative total vdW
energy than 3 well-separated bundles of 2 strands ( i.e. an odd-number
preference).  We also discuss a bundle of 6 strands and explore the
relative contributions beyond pairwise interactions.

\end{abstract}

\newpage
\input{txt/Jeet.tex}

\section*{Acknowledgments}
  SP acknowledges support from the Marie Sk{\l}odowska-Curie Doctoral Network TIMES, grant No. 101118915.  



 





\end{document}

%% file: txt/Jeet.tex
\section*{Introduction}

Highly elongated structures (filaments, strands, wires etc.,) are of widespread
importance in nano-science, chemistry and biology.  Examples include nano-tubes\cite{dresselhaus1998physical}, nano-wires\cite{garnett2019introduction}, polymers, and also common biological structures such as DNA, muscle fibers, and  the glycocalyx layer surrounding living cells\cite{2strandedDNA}. Many of these structures are good electrical conductors\cite{dresselhaus1998physical,jiménez2017high}. Parallel
quasi-one-dimensional metals are known\cite{chang1971van,langbein1972van,mitchell1973van1,dobson2023mbd+,dresselhaus1998physical,PhysRevB.68.205409,drummond2007van} to experience strong
dispersion (vdW) interactions that fall off unusually slowly
with separation between the strands. These forces cause them to coalesce into many-strand
bundles. At the final equilibrium configuration, strong covalent, metallic, or Pauli repulsion forces—as well as hydrogen bonds—may determine the detailed atomic geometry. For most inter-strand separations, however, the vdW force
is dominant and will determine  the initial clustering tendencies. In
these clusters there are potentially strong multi-strand vdW interactions that
go beyond a sum of negative~(attractive) pairwise inter-strand energies. A perturbative analysis~\cite{davies1973van, pal2025attractive, richmond1972many} showed that the additional 3-strand
energy term is positive (i.e., repulsive). This led to the intriguing
speculation~\cite{mitchell1973van1,mitchell1973van2, mitchell1973van3, smith1973van,pal2025attractive,ninham2023nafion} that the strands may prefer to cluster into even-numbered
bundles, which could have implications for the formation kinetics of DNA,
for example. We subsequently showed~\cite{pal2025attractive} that the above trend continues
beyond the triplet level: the leading perturbative $n$-strand vdW energy
correction has a sign of $(-1)^{n-1}$. Thus odd-$n$ corrections are
repulsive and even-$n$ corrections are attractive. We also noted that this sign
alteration implies slow convergence of the perturbative approach.\\
\noindent In order to address the above conjecture about even-cluster preference, we
therefore analyze the vdW energy of 1D metal clusters via a non-perturbative
approach.
\subsubsection*{The test system studying here:} As a concrete example we compute the total vdW energy among six
parallel strands, considering the following spatial configurations: 
\begin{itemize}
    \item three well-separated clusters, each containing two strands   (``$3\times 2$'') [See Fig.\,(\ref{fig:4a})]
    \item  two well separated clusters of  three strands (``$2 \times 3$'') [See Fig.\,(\ref{fig:4b})]
    \item one cluster of six strands in a hexagonal close packed structure (``$1 \times 6$'') [See Fig.\,(\ref{fig:4c})]
\end{itemize}
The near-neighbor distance $R$ is same for all three cases. In these
high-symmetry cases we derive closed analytic expressions for the plasmon
frequencies. We find that the total non-perturbative vdW energies per unit
length are in the order $\rm E^{(1 \times 6)} < E^{(2 \times 3)} < E^{(3 \times 2)} < 0 $. Thus $2 \times 3$ (an odd-numbered clustering) is more bound than $3 \times 2$ (an even-numbered clustering). Thus the question posed in the \textit{title} is
answered in the negative: with a given number (here 6) of stands available,
there is no universal preference for even-$n$ clusters. We also investigate
trends in the role of pairwise vs. beyond-pairwise contributions. \

\section*{Method}

Our strategy for non-perturbative vdW energy calculations involves computing
the frequencies $\omega _{j}$ of the multi-strand coupled plasmons (electron
density oscillations) using the time-dependent Hartree approximation for the
inter-strand Coulomb interaction. At $T=0K$ the vdW energy is then
contained in the sum of plasmon zero-point energies: $\sum_{j}\hbar \omega _{j}/2$. This amounts to treating the inter-strand
correlation energy in the direct Random Phase Approximation (dRPA)\cite{furche2008developing,PhysRevB.68.205409}.
This energy is not exact but it sums an infinite subset of perturbative
terms represented by ring-diagrams in Feynman perturbation theory. \ It becomes exact in the limit of large
inter-strand separation. Notably, the Lifshitz theory of inter-slab
vdW interactions can be derived in the same framework\cite{Maha,SubhojitPCCP2024}.

\noindent To implement the dRPA approach we need know how the electron number
density $n_{I}\left( \vec{r},t\right) $ on a given strand (labeled $I$) responds to the potential $v_{I}\left( \vec{r},t\right)$ generated by density fluctuations  $n_{J}\left( \vec{r},t\right) $ on the other wires labeled $J$.
The dRPA assumes that this response is linear, so each wire has a density response function 
 $\chi _{I}\left( \vec{r},\vec{r}^{\prime },t-t^{\prime
}\right) $ such that 
\begin{equation}
n_{I}=\chi _{I}\ast v_{I}=\chi _{_{I}}\ast \sum_{J\neq I}w_{IJ}\ast n_{J}\;.
\label{TDH_general}
\end{equation}%
where $w_{IJ}\left( \vec{r}-\vec{r}^{\prime }\right) $ denotes the inter-strand
Coulomb interaction and $``\ast$'' represents space-time convolution. \ 

\noindent In the regime that we are treating, the strands are well separated, allowing us to adopt a continuum quasi-one-dimensional model for \ the response $\chi _{I}
$. \ The continuum approximation ignores the atomic graininess of each
strand, except for the introduction of an effective Bloch electron mass $%
m^{\ast }$ to summarize the effects of \ a periodic 1D potential on the
conduction electrons. The quasi-1D approximations incorporates only the axial
motions of \ the conduction electrons and ignores their smaller
polarizability in the directions perpendicular to the strand's long axis. \
It also neglects the polarizability\ arising from non-conduction electrons. These neglected vdW energy contributions can be included efficiently via the MBD+C approach~\cite{dobson2023mbd+} which shows that they are rapidly eclipsed by the axial conduction electron contribution treated here, as the strands are separated away from contact.

\subsection*{Independent-electron response $\protect\chi $ of a single
quasi-1D conducting wire}

In the quasi-1D continuum model, the the density response of a strand is
translationally invariant along the $z$ direction (parallel to the long axis):
\begin{equation}
\begin{aligned}
\chi _{I}(\vec{r},\vec{r}^{\prime },t-t^{\prime }) &=\rho \left( \vec{r}%
_{\perp }\right) \rho \left( \vec{r}_{\perp }^{\prime }\right) \chi \left(
z-z^{\prime },t-t^{\prime }\right)\\
&=\left( 2\pi \right)^{-2}\rho \left( \vec{r}_{\perp }\right) \rho \left( 
\vec{r}_{\perp }^{\prime }\right) \int d\omega~dq~\chi\left( q,\omega \right)~e^{i\left[(z-z^{\prime
})-\omega \left( t-t^{\prime }\right) \right]} 
\end{aligned}
\label{1D_contiuum_chi} 
\end{equation}%
Here $q\equiv q_{z}$ and cylindrical coordinates $\ \vec{r}=\vec{r}_{\perp }+
$ $z\hat{z}\equiv (r_{\perp },\phi ,z)$\ were used. $\rho \left( r_{\perp
}\right) $ is the normalized, angularly averaged ground-state conduction
electron number density of the strand.

\noindent The conduction electron density response $\chi$ of an isolated strand will first be specified for independent electrons, i.e., with neglect of the Coulomb interaction between electron density fluctuations.  The following form for small wave vectors $q$  can be derived by hydrodynamic arguments or by using the time-dependent Schr\"odinger mechanics for Bloch electrons:
\begin{equation*}
\chi _{0}\left( q,\omega \right) =\frac{n_{0}q^{2}}{m^{\ast }\left( \omega
^{2}-\beta ^{2}q^{2}\right) }+ \mathcal{O}\left( q^{6}\right), \quad q\to 0 
\end{equation*}%
Here $m^{\ast}$ is an effective mass for Bloch electrons and $\beta$ is a
characteristic diffusion velocity. $n_0$ is the equilibrium conduction electron number density. $\beta$ is comparable to the Fermi
velocity for the case of degenerate electrons, and comparable to the
thermal velocity for non-degenerate electrons.

\section*{The electron-electron Coulomb interaction between and within
strands}

The Coulomb interaction between well-separated strands is not very sensitive
to the cross sectional shape of the strands. We
model it by assuming that all charges are concentrated at the strand's
central axis, so that the inter-strand Coulomb energy is 
\begin{equation*}
V(z-z^{\prime },D)=\frac{\left| e\right| ^{2}}{\sqrt{\left( z-z^{\prime
}\right) ^{2}+D^{2}}}
\end{equation*}%
where $D$ is the distance between the central axes of the parallel strands.
The 1D Fourier transform is:%
\begin{equation}
w^{\rm inter}= V\left( q,D\right) =\int_{-\infty }^{\infty }\frac{\left| e\right| ^{2}}{%
\sqrt{Z^{2}+D^{2}}}e^{-iqZ}dZ=2\left| e\right| ^{2}K_{0}\left( \left|
q\right| D\right) 
\label{v_inter}
\end{equation}%
where $K_{0}$ is modified Bessel function of zero order (see Ref.\,\cite{abramowitz1964handbook} formula 9.6.21]).

\noindent The intra-strand Coulomb interaction is more sensitive to the
cross-sectional density profile. For definiteness here we assume a hollow
cylinder (representing a nanotube for example) so that%
\begin{equation}
\rho \left( \vec{r}_{\perp }\right) =\left( 2\pi a\right) ^{-1}\delta \left(
r_{\perp }-a\right)   \label{Nanotube_radial_density}
\end{equation}%
where $a$ is the radius of the cylinder. \ Other \ choices of $\rho \left( 
\vec{r}_{\perp }\right) $ yield the same results as $q\rightarrow 0$, a limit that describes vdW interactions between well-separated
strands.

The response from Eqs.~(\ref{1D_contiuum_chi}) and (\ref{Nanotube_radial_density}%
) describes the motion along the $z$ axis of rigid
rings of charge. The Coulomb energy between two rings is 
\begin{eqnarray}
\widetilde{V}(z-z^{\prime }) &=&\int d^{2}r_{\perp }\int d^{2}r_{\perp }^{\prime
}\frac{e^{2}}{\sqrt{(z-z^{\prime })^{2}+\left| \vec{r}_{\perp }-\vec{r}'%
_{\perp }\right| ^{2}}}\rho \left( \vec{r}_{\perp }\right)  \rho (\vec{r}%
_{\perp }^{\prime })  \notag \\
&=&\frac{1}{(2\pi )^{3}}\int d^{3}q\int d^{2}r_{\perp }\int d^{2}r_{\perp
}^{\prime }\frac{4\pi e^{2}}{q_{||}^{2}+q_{\perp }^{2}}~e^{i\vec{q}_{\perp }\cdot (%
\vec{r}_{\perp }-\vec{r}'_{\perp})}~e^{iq_{||}(z-z^{\prime })}\rho (\vec{r}_{\perp })%
\rho(\vec{r}_{\perp }^{\prime })  \notag \\
\widetilde{V}(q) &=&\frac{1}{(2\pi )^{2}}\int d^{2}q_{\perp }\frac{4\pi e^{2}}{%
q_{||}^{2}+q_{\perp }^{2}}\bar{\rho}\left( \vec{q}_{\perp }\right)\bar{\rho%
}(-\vec{q}_{\perp })  \label{w_intr_from_rho(qperp)}
\end{eqnarray}

\noindent Here for the hollow cylinder model
\begin{equation*}
\begin{aligned}
\bar{\rho}(q_{\perp })&=\int_{-\pi }^{\pi }d\phi \int_{0}^{\infty
}dr_{\perp }r_{\perp }\frac{\delta(r_{\perp }-a)}{2\pi a}~e^{iq_{\perp
}r_{\perp }\cos \phi } \\
&=\frac{1}{2\pi }\int_{-\pi }^{\pi }d\phi~e^{iq_{\perp}a\cos\phi
}=J_{0}(q_{\perp }a)
\end{aligned}
\end{equation*}%
where $J_{0}$ is the Bessel function of zeroth order. Then, Eq.~(\ref{w_intr_from_rho(qperp)}) gives the intra-strand Coulomb
interaction as%
\begin{equation}
w^{\rm intra}(q)=\frac{1}{(2\pi )^{2}}\int_{0}^{\infty }2\pi~dq_{\perp }q_{\perp }\frac{%
4\pi \left| e\right| ^{2}}{q_{||}^{2}+q_{\perp }^{2}}\left(
J_{0}(\left| q_{\perp }\right| a)\right) ^{2}=2\left| e\right|
^{2}I_{0}(a\left| q\right|) K_{0}(a\left| q\right| )
\label{intra}
\end{equation}%
Here the integral is taken from formula 6.541.1 of Ref.\,\cite{gradshteyn2014table}.

\section*{The density response of a single strand in the dRPA}

The \ dRPA equation for the density perturbation $n\left(q,\omega \right)
\rho \left( r_{\perp }\right) \exp\left( iqz-i\omega t\right) $
due to an external potential $v=v(q,\omega) \exp\left(
iqz-i\omega t\right) $ is

\begin{equation*}
n=\chi _{0}\left( v+w^{\rm intra}~n\right) \implies n=\frac{\chi _{0}}{%
1-w^{\rm intra}~\chi_{0}}~v
\end{equation*}%
where $w^{\rm intra}$ is given by Eq.\,(\ref{intra}). Thus%
\begin{equation}
\begin{aligned}
\chi^{\mathrm{dRPA}}(q, \omega) = \frac{n}{v}=\frac{\chi_{0}}{%
1-w^{\rm intra}~\chi_{0}} =  \frac{n_{0}q^{2}}{m^{\ast }\left( \omega^{2}-\Omega
_{1D}^{2}\left( q\right) \right) }
\end{aligned}
\label{chi_drpa}
\end{equation}%
Here $\Omega _{1D}=c_{1D}~|q|\sqrt{\left(
I_{0}\left( qa\right) K_{0}\left( qa\right) +\frac{\beta ^{2}}{c_{1D}^{2}}%
\right)}$ is the 1D plasmon frequency and $c_{1D}=\sqrt{%
2n_{0}\left| e\right| ^{2}/m^{\ast }}$ is the characteristic 1D velocity.

\section{The coupled plasmon frequencies and vdW energy of two parallel
strands}
 \begin{figure}[!ht]
        \centering
        \includegraphics[width=0.2\linewidth]{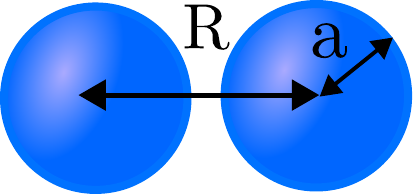}
        \caption{(Colors online) A single pair of strands.}
        \label{fig:two}
    \end{figure}
Within the \ continuum approximation [Eq.\,(\ref{1D_contiuum_chi})], the space and
time convolutions in Eq.\,(\ref{TDH_general}) reduce to simple products in $(q,\omega)$
space. For two parallel identical conducting wires with separation $R$ shown in Fig.\,(\ref{fig:two}), the governing dRPA equations for free coupled oscillations of the electron number density
perturbations \ $n_{1}\exp( iqz-i\omega t)$ and $%
    n_{2}\exp( iqz-i\omega t)$ take a symmetric form:
\begin{equation*}
        n_{1} =\chi w^{\rm inter} n_{2}, \quad n_{2} =\chi w^{\rm inter} n_{1}
    \end{equation*}%
    where $w^{\rm inter}$ is given by Eq.\,(\ref{v_inter}) and $\chi$ is the
    single-wire response \ given by Eq.\,(\ref{chi_drpa}).
This requires $(\chi\,w^{\rm inter})^2 = 1\implies \chi\,w^{\rm inter} = \pm 1$, which in turn gives rise to two branches of plasmon frequencies:
    \begin{equation}
    \begin{aligned}
        \omega_{\pm}^{2} &=\Omega^{2}\left(q\right) \pm c_{1D}^{2}q^{2}K_{0}\left(
        qR\right)  \\ 
        &= c^{2}_{1D}q^{2}\left(
        K_{0}\left( qb\right) I_{0}\left( qb\right) +\gamma \pm K_{0}\left( qR\right)
        \right)
        \end{aligned}
    \end{equation}%
    where $\gamma \equiv \beta^{2}/c_{1D}^{2}$ is the ratio of kinetic pressure to
    Coulomb pressure in a wire. The zero-temperature vdW interaction energy per unit length is obtained by subtracting the plasmon zero-point energies:
   
\begin{eqnarray*}
     \rm{ E_{\mathrm{vdW}}/L} =\frac{\hbar}{2}\frac{1}{2\pi }2\int_{0}^{\infty }\Big[\left(\omega_{+}\left(
        \left| q\right|, R\right) +\omega_{-}\left( \left| q\right| ,R\right) \right)-
        2\omega_{+}\left( q,\infty \right)\Big] dq \\ =\frac{\hbar c_{1D}}{2\pi }\int_{0}^{\infty }
        dq\,q \Big(\sqrt{K_{0}\left( aq\right) I_{0}\left( aq\right) +\gamma + K_{0}\left(
        Rq\right) }+\sqrt{%
        K_{0}\left( aq\right) I_{0}\left( aq\right) +\gamma -K_{0}\left(Rq\right)
        } \\ -2\sqrt{K_{0}\left( aq\right) I_{0}\left( bq\right) +\gamma }\Big)
    \end{eqnarray*}%
    \[
        \rm{E_{\mathrm{vdW}}(\rho)/L}=\frac{\hbar c_{1D}}{2\pi a^{2}}\varepsilon
        ^{\left( 2\right) }\Big(\rho,\gamma\Big)
    \]
    where
    \begin{equation}
        \begin{aligned}
            \varepsilon^{\left( 2\right) }\left(\rho,\gamma \right) =\int_{0}^{\infty }dQ Q \Big( \sqrt{K_{0}\left( Q\right) I_{0}\left( Q\right) +\gamma +K_{0}\left( Q \rho\right) }+\sqrt{K_{0}\left(Q\right) I_{0}\left( Q\right) +\gamma -K_{0}\left( Q\rho\right) } \\
            -2\sqrt{K_{0}\left( Q\right) I_{0}\left( Q\right) +\gamma}\Big),\;\;Q\equiv qa\,\,\text{and}\,\, \rho = R/a
        \end{aligned}
        \label{E2}
    \end{equation}%


\section*{The coupled plasmon frequencies and vdW energy of three parallel
strands}

\begin{figure}[!h]
        \centering
        \includegraphics[width=0.23\linewidth]{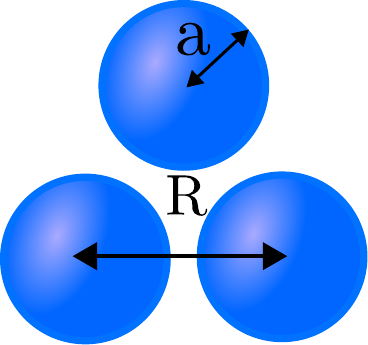}
        \caption{(Colors online) Three parallel strands in an equilateral triangular geometry.}
        \label{fig:three}
    \end{figure}

    Similarly the dRPA equations for the coupled plasmon eigenmodes for three identical parallel conducting wires (placed at the vertices of an equilateral triangle with side $R$ in Fig.\,(\ref{fig:three})) are%

    \begin{equation}
        \begin{aligned}
            n_{1}=\chi w^{\rm inter} \left( n_{2}+n_{3}\right)                           \\
            n_{2}=\chi w^{\rm inter} \left( n_{1}+n_{3}\right)                           \\
            n_{3}=\chi w^{\rm inter} \left( n_{1}+n_{2}\right) 
        \end{aligned}
        \label{3object_R PA_eqns}
    \end{equation}
where $w^{\rm inter}$ is given by Eq.\,(\ref{v_inter}). The discrete angular
    translational symmetry requires that the (un-normalized) eigenfunction
    $\vec{n}^{\left( J\right) }$ have Bloch-like form (eigenfunctions of
    the discrete rotation ($2\pi/3$ rad) operator):
    \begin{equation}
        n_{I}^{\left( J\right) }=e^{2\pi I J/3};\;J=0,1,2 \label{Rotational_eigemfunctions}
    \end{equation}%
    where $I$ labels the three wires and $J$ denotes the three coupled plasmon
    modes. Substituting Eq.\,(\ref{Rotational_eigemfunctions}) into any one of the
    three equations in  Eq.\,(\ref{3object_R PA_eqns}) yields $\chi \left( q,\omega _{0}\right)~w = 1/2$ for $J=0$ and $\chi \left( q,\omega \right)~w = - 1$ for $J = 1, 2$.
    Using Eq.\,(\ref{chi_drpa}) for $\chi \left( q,\omega \right)$ and
    Eq.\,(\ref%
    {v_inter}) for $w_{q}$, we find the coupled mode frequencies $%
    \omega_{J}$:
    \begin{equation*}
    \begin{aligned}
        \omega_{J=0}^{2}\equiv \omega_{0}^{2}=c_{1D}^{2}q^{2}\left( K_{0}\left( a
        \left| q\right| \right) I_{0}\left( a|q|\right) +\frac{\beta^{2}}{c_{1D}^{2}}
        +2K_{0}\left( |q|R\right) \right) \\
          \omega_{1}^{2}=\omega_{2}^{2}=c_{1D}^{2}q^{2}\left( K_{0}\left( a\left| q
        \right| \right) I_{0}\left( a\left| q\right| \right) +\frac{\beta^{2}}{%
        c_{1D}^{2}}-K_{0}\left( \left| q\right| R\right) \right)
        \end{aligned}
    \end{equation*}%
\noindent The vdW energy per unit length is then
    \begin{eqnarray*}
        \rm{E_{vdW}/L} =\frac{1}{2\pi }\frac{\hbar }{2}\int_{-\infty }^{\infty }%
        \left[ \omega _{0}\left( \left| q\right| ,R\right) +2\omega _{1}\left(
        \left| q\right|, R\right) -\omega _{0}\left( \left| q\right| ,\infty
        \right) -2\omega_{1}\left( \left| q\right| ,\infty \right) \right] dq \\
        =\frac{1}{2\pi }\frac{\hbar }{2}c_{1D}\int_{-\infty }^{\infty } dq \left|
        q\right| \Bigg[
        \sqrt{K_{0}\left( a\left| q\right| \right) I_{0}\left( a\left| q\right| \right)
        +\frac{\beta^{2}}{c_{1D}^{2}}+2K_{0}\left( \left| q\right| R\right) }\\ -3\sqrt{K_{0}\left(
        a\left| q\right| \right) I_{0}\left( a\left| q\right| \right) +\frac{\beta
        ^{2}}{c_{1D}^{2}}} + 2\sqrt{K_{0}\left( a\left| q\right| \right) I_{0}\left(
        a\left| q\right| \right) +\frac{\beta^{2}}{c_{1D}^{2}}-K_{0}\left( \left|
        q\right| R\right)}\, \Bigg]
    \end{eqnarray*}

    \noindent This can be written%
    \begin{equation*}
        \rm{E_{vdW}(\rho)/L}=\frac{\hbar c_{1D}}{2\pi a^{2}\;}\varepsilon^{\left(
        3\right) }\left( \rho, \gamma\right)
    \end{equation*}%
    where the dimensionless total 3-object vdW energy is%
    \begin{equation}
        \begin{aligned}
            \varepsilon^{\left( 3\right) }\left(\rho,\gamma \right) =\int_{0}^{\infty }dQ Q\Big[ \sqrt{K_{0}\left( Q\right) I_{0}\left( Q\right) +\gamma +2K_{0}\left( Q\rho\right) }+2\sqrt{K_{0}\left( Q\right) I_{0}\left(Q\right) +\gamma -K_{0}\left( Q\rho\right)} \\
            -3\sqrt{K_{0}\left(Q\right) I_{0}\left( Q\right) +\gamma }\Big] ,\;\;Q\equiv qa\,\,\text{and}\,\, \rho = R/a
        \end{aligned}
        \label{E3}
    \end{equation}



\section{The coupled plasmon frequencies and vdW energy of six parallel
strands}

We evaluate the vdW energy of the most compact and symmetric array of \ six
strands, which is a section of a hexagonal close packed lattice shown in Fig.\,(\ref{fig:six}). The distance between the centers of neighboring strands is $R$, where the characteristic inter-strands distances are: $R$ (nearest neighbors), $2R$ and $2 R \cos \left(30^{\circ}\right)=$ $\sqrt{3} R$ (next-to-nearest neighbors). The strands form two sub-lattices grouped by triangles: corner wires (1,3,5) and mid-edge wires (2,4,6), as illustrated in Fig.\,(\ref{fig:six}). 
The system is symmetric under $120^{0}$ rotations. It implies that the plasmon normal modes must constitute a complete set of eigenfunctions for the discrete rotational operator. The electron number density perturbation for the corner wires will have the same form as: 

\begin{figure}[!h]
    \centering
    \includegraphics[width=0.3\linewidth]{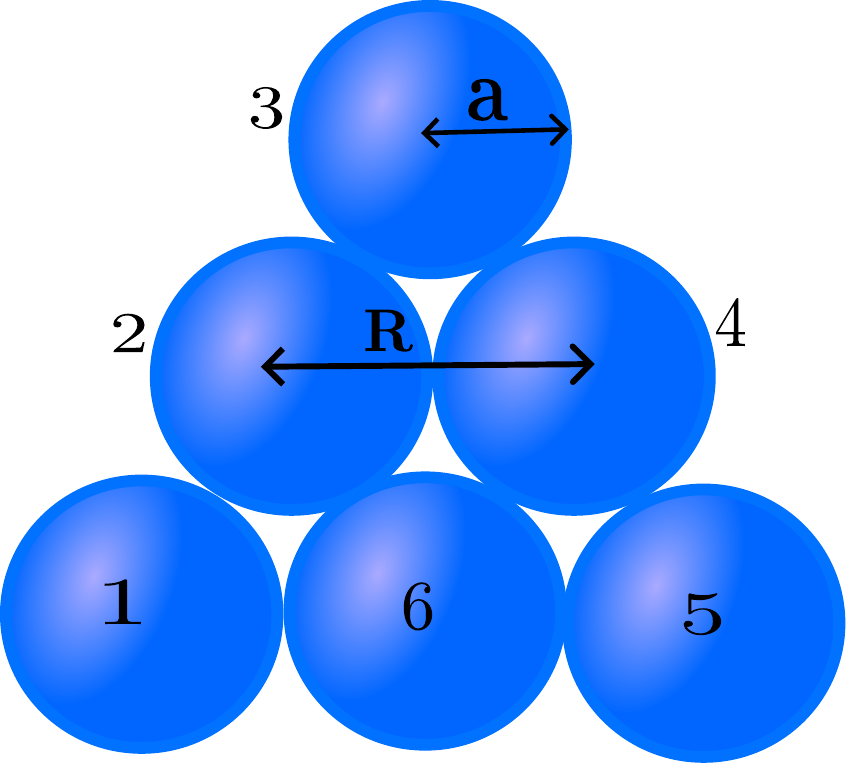}
    \caption{(Colors online) Six strands in hexagonal close packed configuration}
    \label{fig:six}
\end{figure}

\begin{equation}
n_{2 j+1}^{(\mu)}=\alpha \exp (2 \pi i \mu j / 3) \exp (i q x-i \omega t), \quad j=0,1,2 \quad \mu=0, 1, 2 \label{10}
\end{equation}

\noindent and the other strands labeled $2,4,6$ we have
\begin{equation}
n_{2 j+2}^{(\mu)}=\beta \exp (2 \pi i \mu j / 3) \exp (i q x-i \omega t), \quad j=0,1,2\,\quad \mu=0,1, 2 \label{11}
\end{equation}

\noindent Here $j$ denotes the wires within each sub-system, and $\mu$ labels the collective modes.
We will find a $2 \times 2$ eigenvalue equation linking to the amplitudes $\alpha, \beta$ for each value of mode $\mu$, giving two modes. $\mu \pm$, for each $\mu$, leading to six modes as expected for six wires. The dRPA equations for wires 1 and 2 are as follows:(Because of the form of the above ansatz Eq.~(\ref{10}, \ref{11}),  these 2 equations follow if we make any one even and one odd filaments as the subjects on the LHS of the RPA equation).


\begin{equation}
    \begin{aligned}
      \alpha= &2 \chi w_{2 R} \cos (2 \pi \mu / 3) \alpha+\chi\left(2 w_{R} \cos (2 \pi \mu / 3)+w_{\sqrt{3} R}\right) \delta
    \end{aligned}   
    \label{12}
\end{equation}
and 
\begin{equation}
    \begin{aligned}
 \delta = & \chi\left(2 w_{R} \cos (2 \pi \mu / 3)+w_{\sqrt{3} R}\right) \alpha + 2 \chi w_{R} \cos (2 \pi \mu / 3) \delta
    \end{aligned}
    \label{13}
\end{equation}


\noindent where $w_{r}=2|e|^{2} K_{0}(q r)$ in Eq.\,(\ref%
    {v_inter}) and $\delta \equiv  \beta \exp (2 \pi i \mu / 3)$ 





\noindent The eigenvalue problem takes the matrix form from Eqs.\,(\ref{12}) and (\ref{13}):
$$
\left(\begin{array}{cc}
\omega^{2}-\Omega(q)^{2}-\omega_{A}^{2}(q, \mu) & -\omega_{B}^{2}(q, \mu) \\
-\omega_{B}^{2}(q, \mu) & \omega^{2}-\Omega^{2}(q)^{2}-\omega_{C}^{2}(q, \mu)
\end{array}\right)\begin{pmatrix}
    \alpha \\
    \delta
\end{pmatrix}=\begin{pmatrix}
    0 \\0
\end{pmatrix}
$$



\noindent with non-zero plasmon modes

\begin{equation}
\begin{aligned}
\left(\omega^{(\mu \pm)}\right)^{2} & =\lambda=\Omega^{2}+\frac{1}{2} \omega_{A}^{2}+\frac{1}{2} \omega_{C}^{2}+\frac{1}{2} \sqrt{\left(\omega_{A}^{4}-2 \omega_{A}^{2} \omega_{C}^{2}+\omega_{C}^{4}+4 \omega_{B}^{4}\right)} \\
& =\Omega^{2}+\frac{1}{2}\left(\omega_{A}^{2}+\omega_{C}^{2}\right) \pm \frac{1}{2} \sqrt{\left(\left(\omega_{A}^{2}-\omega_{C}^{2}\right)^2+ 4 \omega_{B}^{4}\right)} 
\end{aligned}
\label{14}
\end{equation}
where the coupling terms are 
$$
\begin{aligned}
\omega_{A}^{2} & =c_{1 D}^{2} q^{2} 2 K_{0}(2|q| R) \cos (2 \pi \mu / 3) \\
\omega_{B}^{2} & =c_{1 D}^{2} q^{2}\left(2 \cos (2 \pi \mu / 3) K_{0}(|q| R)+K_{0}(\sqrt{3}|q| R)\right) \\
\omega_{C}^{2} & =c_{1 D}^{2} q^{2} 2 \cos (2 \pi \mu / 3) K_{0}(|q| R)
\end{aligned}
$$


\noindent We now define  dimensionless frequencies
$$
\zeta: \omega^{(\mu \pm)}(q)=c_{1 D} a^{-1}|Q| \zeta^{m \pm}, \quad Q \equiv a q
$$

\noindent and a dimensionless spatial separation variable $\rho=R / a$. Then Eq.\,(\ref{14}) becomes

$$
\begin{aligned}
\zeta^{(\mu \pm)}(q)^{2} & =\zeta_{1 D}^{2}+\frac{1}{2}\left(\zeta_{A}^{2}+\zeta_{C}^{2}\right) \pm \frac{1}{2} \sqrt{\left(\zeta_{A}^{2}-\zeta_{C}^{2}\right)^{2}+4 \zeta_{B}^{4}} \\
\zeta_{1 D}^{2} & =K_{0}(Q) I_{0}(Q)+ \gamma \\
\zeta_{A}^{2} & =2 \cos (2 \pi \mu / 3) K_{0}(2 \rho Q) \\
\zeta_{B}^{2} & =2 \cos (2 \pi \mu / 3) K_{0}(\rho Q)+K_{0}(\sqrt{3} \rho Q) \\
\zeta_{C}^{2} & =2 \cos (2 \pi \mu / 3) \quad \rho=R / a, \quad Q \equiv q a
\end{aligned}
$$

\noindent For the $\mu=0$ modes :

$$
\begin{aligned}
\zeta_{A}^{2} & =2 K_{0}(2 \rho|Q|) \\
\zeta_{B}^{2} & =2 K_{0}(\rho|Q|)+K_{0}(\sqrt{3} \rho|Q|) \\
\nu_{C}^{2} & =2 K_{0}(\rho|Q|)
\end{aligned}
$$

\noindent For $\mu=1$ and $\mu=2$ (degenerate, just as for 3 wires in equilateral configuration): Using $2 \cos (2 \pi / 3)=-1, \quad 2 \cos (4 \pi / 3)=-1$ we have

$$
\begin{aligned}
\zeta_{A}^{2} & =-K_{0}(2 \rho|Q|) \\
\zeta_{B}^{2} & =-K_{0}(\rho|Q|)+K_{0}(\sqrt{3} \rho|Q|) \\
\zeta_{C}^{2} & =-K_{0}(\rho|Q|)
\end{aligned}
$$

\noindent Note: some of these dimensionless frequencies $\zeta$ are imaginary, but only their squares appear in the expressions for the mode frequencies. Also note that the modes for $\mu=1$ and $\mu=2$ are degenerate.

\noindent The dispersion energy per unit length between the wires at $T=0 K$ 
\begin{equation}
\begin{aligned}
\rm E_{\rm{vdW}}(\rho) / L= & \frac{\hbar}{2} \frac{1}{2 \pi} \int_{-\infty}^{\infty} a^{-1} d Q c_{1 D} a^{-1}|Q| \left(\zeta^{(0+)}+\zeta^{(0-)}+2 \zeta^{(1+-)}+2 \zeta^{(1-)}-6 \zeta_{1 D}\right) \\
= & \frac{\hbar c_{1D}}{4 \pi a^{2}} \int_{-\infty}^{\infty}\left(\zeta^{(0+)}+\zeta^{(0-)}+2 \zeta^{(1+-)}+2 \zeta^{(1-)}-6 \zeta_{1 D}\right)|Q| d Q \\
= & \frac{\hbar c_{1D}}{2 \pi a^{2}} \int_{0}^{\infty}\left(\zeta^{(0+)}+\zeta^{(0-)}+2 \zeta^{(1+-)}+2 \zeta^{(1-)}-6 \zeta_{1 D}\right) Q d Q
\end{aligned}
\end{equation}

\noindent The integral is a dimensionless function of the dimensionless separation variable $\rho=R / a$ where $a$ is the radius of each wire.

\section*{Testing the even-clustering conjecture}

We aim to test the intriguing conjecture~\cite{mitchell1973van1,mitchell1973van2, mitchell1973van3, smith1973van,pal2025attractive,ninham2023nafion} that there is an
universal tendency for parallel metallic strands to aggregate into
even-numbered clusters rather than odd-numbered ones. A single
counterexample will be enough to invalidate this conjectures as a general
proposition. \ \ The original conjecture was based on perturbative assessment of the
triplet term, but perturbative convergence is slow for three strands, so
we investigate this using our non-perturbative results above. 



\begin{figure}[!htbp]
    \centering
    \begin{subfigure}[t]{0.35\textwidth}
        \centering
        \includegraphics[width=\linewidth]{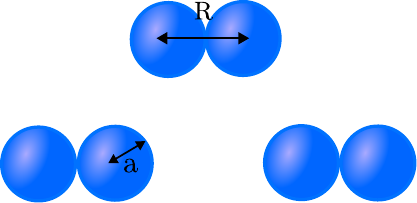}
        \caption{}
        \label{fig:4a}
    \end{subfigure}%
    \hspace{1in}
    \begin{subfigure}[t]{0.32\textwidth}
        \centering
        \includegraphics[width=\linewidth]{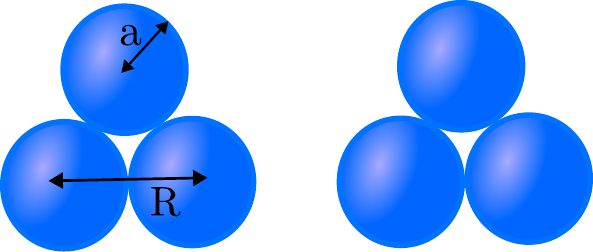}
        \caption{}
        \label{fig:4b}
    \end{subfigure}

    \vspace{0.3em} 

    \begin{subfigure}[t]{0.21\textwidth}
        \centering
        \includegraphics[width=\linewidth]{six_close.pdf}
        \caption{}
        \label{fig:4c}
    \end{subfigure}

    \caption{(a) Three well separated identical clusters of two strands with radius $a$ and inter-strand separation $R$ (b) Two well separated identical clusters of three strands with radius $a$ and inter-strand separation $R$ placed in the vertices of a equilateral triangle. (c) A cluster of six parallel  identical strands with radius $a$ and inter-strand separation $R, \sqrt{3}R$ and $2R$ with nearest neighbor and next-to-nearest neighbors placed in a hexagonal closed pack structure. }
    \label{fig:foobar}
\end{figure}

\noindent We consider six strands in Fig.\,(\ref{fig:4c}). We will first compare the total vdW energies of two
configurations :\ (a) three widely separated clusters, each containing two
nearby strands [``$3 \times 2$'', Fig.\,(\ref{fig:4a})]; and (b) two well separated clusters,
each consisting of \ thee nearby strands [''$2\times 3"$, Fig.\,(\ref{fig:4b})].

\begin{figure}[!h]
    \centering
    \includegraphics[width=0.8\linewidth]{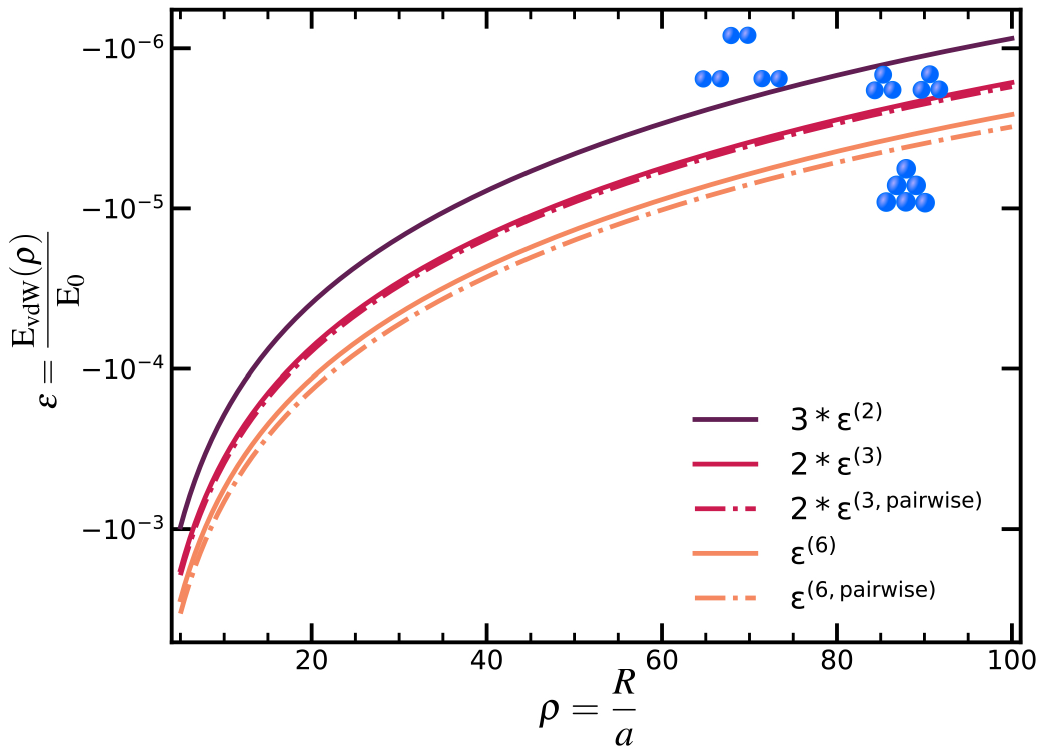}
    \caption{Colors online: Dimensionless vdW energy per unit length $ \mathrm{{ \varepsilon}} = \rm{E_{vdW}}(\rho)/\rm{E}_0$ v/s dimensionless inter-filament separation $\rho = \rm{R}/\rm{a}$. When the filaments are (5,5) CNTs, $\rm{E}_{0}= $ 0.13 eV/\text{\AA} and $ \rm{a} = 3.5 \text{\AA}$.  Solid curves: non-perturbative energies. Dashed curves:  sum of pairwise energies.    Black curve: 3 widely-separated clusters of  2 filaments.  Red curves: 2 widely-separated clusters of 3 filaments. Orange curves: single cluster of 6 filaments.}
    \label{fig:result}
\end{figure}

\noindent The vdW energies of \ these two configurations are as follows: 
\begin{equation*}
\varepsilon ^{\left( 3\times 2\right) }=3\varepsilon ^{^{\left( 2\right)
}},\;\;\varepsilon ^{\left( 2\times 3\right) }=2\varepsilon ^{\left(
3\right) }\;\;.
\end{equation*}%
In Fig.\,(\ref{fig:result}) (solid lines) the energies $\varepsilon ^{\left( 3\times
2\right) }$ and $\varepsilon^{\left( 2\times 3\right)}$ are
plotted in a dimensionless \ form versus dimensionless intra-cluster strand
separation $R/a$. The plot reveals that the $2\times 3$ configuration is
more bound than the $3\times 2$ configuration at all intra-cluster
separations. \ This is a counterexample to the conjecture of \ an
even-cluster preference, since the energetically favored $2\times 3$
configuration consists exclusively of odd-numbered clusters while the less favorable arrangement $3 \times 2$ comprises only even-numbered clusters. So the question posed in the Title is answered in the
negative: there is no general preference of an even-numbered clusters. 

\section*{Pairwise theory as a predictor of cluster  preference}

In Fig.\,(\ref{fig:result}), we also show \ the energies based on pairwise summation of
two-strand energies (dashed lines). \ The full theory reveals that the
odd-cluster $2\times 3$ configuration is slightly less stab than the pairwise
estimate, as suggested by the original observation from the repulsive triplet vdW
energy contributions. However the beyond-pairwise
contributions evidently are not large repulsive enough to reverse the energy
ordering predicted by the pairwise approximation. \ 

\noindent In general, the pairwise theory predicts that the preference is not based on odd-
even considerations: instead, the energetically favored configurations are those
that have the maximum number of strong near-neighbor (nn) pairs of strands.
Configuration $3\times 2$ has 3 nn bonds, while $2\times 3$ has 6 nn bonds and so is favored over $3\times 2$.


Pairwise theory also suggests that even lower energies can be obtained by bringing more of the 6 strands together. The single 6-strand cluster (``$1 \times 6$'') [see Fig.\,(\ref{fig:4c})] has 9 nn pairs and so should be more strongly bound than either $3 \times 2$ or $2 \times 3$ configurations. See the dashed orange line in Fig.\,(\ref{fig:result}). The full
non-perturbative calculation (solid orange line) confirms that $1\times 6$ is
the most energetically favored among these three configurations studied in this letter. The beyond-pairwise effects in the full theory do somewhat reduce the energetic advantage of $1 \times 6$, but not enough to change the energy ordering predicted by pairwise theory.

\section*{Summary}

It has previously been conjectured~\cite{mitchell1973van1,mitchell1973van2, mitchell1973van3, smith1973van,pal2025attractive,ninham2023nafion} that \ parallel metallic
strands prefer to form even-numbered vdW \ clusters, which could have
significant consequences in biology and nano-science. Here we investigated this for
the case of 6 strands in various geometric configurations. We used a
quasi---one dimensional continuum model for the \ electronic response of each
strand, which allowed us to obtain closed analytic form of vdW energy for the coupled
plasmon frequencies in a non-perturbative analysis via plasmon zero-point energy summation.  We found that two
widely-separated clusters of three strands are more bound \ than three
widely-separated \ clusters of \ two strands. This amounts to a preference for \emph{odd}-numbered clustering, so the above conjecture cannot be
universally true. \ We further found that a single cluster of six strands
has the lowest energy, not because of its even-numberedness for constructing the maximal number of strong nearest-neighbor van der Waals bonds. Surprisingly, simple
pairwise summation correctly predicts the energy ordering of the various
geometric configurations, despite the potentially rich physics of the
beyond-pairwise vdW interactions in theses highly polarizable systems.